\documentclass[fleqn,twoside]{article}
\usepackage{espcrc2}

\usepackage{graphicx}
\usepackage[figuresright]{rotating}

\newcommand{\AmS}{{\protect\the\textfont2
  A\kern-.1667em\lower.5ex\hbox{M}\kern-.125emS}}

\title{Topological Charge Membranes in 2D and 4D Gauge Theory\thanks{Talk 
presented by H. Thacker}}
\author{
        H. Thacker\address[UVA]{Dept. of Physics, University of Virginia, Charlottesville, VA 22904},
        S.~Ahmad\addressmark[UVA], 
        J.~Lenaghan\addressmark[UVA],
       }
       
\begin{document}

\begin{abstract}
Local topological charge structure in the 2D CP(N-1) 
sigma models is studied using the overlap Dirac operator.
Long-range coherence of topological charge along locally 1D regions in 2D space-time is observed. 
We discuss the connection between these results
and the recent discovery 
of coherent 3D sheets of topological charge in 4D QCD. In both 
cases, coherent regions of topological charge form along 
surfaces of approximmate codimension 1.
\vspace{1pc}
\end{abstract}

\maketitle

\section{Introduction}

A variety of lattice studies of topological susceptibility and of the flavor singlet pseudoscalar hairpin correlator in QCD have 
confirmed the essential theoretical understanding of the role of topology and the axial anomaly in generating a
large mass for the $\eta'$ meson. In spite of these results, the detailed local structure of topological charge
fluctuations and their possible relation to chiral symmetry breaking and quark confinement are not well understood. 
In considering the various possibilities for the structure of topological charge fluctuations, it is important to
keep in mind a fundamental constraint imposed by spectral positivity on the two-point Euclidean topological charge 
correlator $\langle Q(x)Q(0)\rangle$, where, in QCD, $Q(x) = (g^2/48\pi^2)Tr F\tilde{F}$. Namely, this correlator must be 
{\it negative} for any nonzero separation $x$. Thus, for example, dominance of $Q$ fluctuations by an uncorrelated gas of finite-size instantons and anti-instantons is ruled 
out by the fact that this would give a positive correlator over a distance characterized by the instanton diameter.
(Of course instantons may still be present, but their contribution to the correlator must be cancelled, e.g. by an
anti-correlated background of quantum fluctuations.) 

Although the negativity of the correlator rules out the dominance of bulk-coherent topological charge fluctuations, 
it allows a more subtle form of long-range order, in which {\it coherent regions 
of topological charge form along lower-dimensional surfaces}.
Just such behavior has recently been observed \cite{Horvath} 
in pure-glue 4D QCD gauge configurations, using the overlap Dirac operator to
define $Q(x)$ on the lattice. It was observed in this study that the topological charge distribution of a typical
gauge configuration was dominated by two coherent, oppositely charged 3D sheets of charge which are everywhere close together,
forming a sort of crumpled dipole layer. Together, the two sheets ``percolate'' through 4D space,
occupying about $80\%$ of the sites on the lattice
(in a box of approximately (1.5 fm)$^4$). 
The fact that the sheets are very thin in one direction and that 
the two oppositely charged sheets remain close together 
allows the observed long-range coherence to occur without violating the negativity of the $Q(x)$ correlator.
It is plausible to suspect that the thickness of the sheets is related to the pseudoscalar glueball mass, while
the long-range coherence along the sheets is associated with Goldstone boson propagation (via delocalized Dirac
eigenmodes).
 
In the work described here, we try to shed some additional light on 
the long-range structure of topological charge
in gauge theories by studying the two-dimensional $CP(N-1)$ models, 
which are believed to have properties
closely analogous to 4D QCD, particularly with respect to topological 
charge structure. From both large N and lattice
strong-coupling arguments, the picture 
of topological charge structure which emerges in $CP(N-1)$ is similar to that originally discussed by Coleman in the massive
Schwinger model. $\theta$ is interpreted as a background electric field, and
there are multiple, discrete ``k-vacua'' which contain k additional units of background electric flux. Charged
particles play the role of domain walls which separate adjacent k-vacuum states differing by a unit of flux.
The free energy $E(\theta)$ exhibits nonanalytic cusps at $\theta =$ odd multiples of $\pi$, which represent 
the crossing of two k-vacuum energies. As the energies cross, the true vacuum state goes from one k-vacuum to
the next by polarizing a charged pair from the vacuum and sending them to infinity in opposite directions. 

The generalization of this k-vacuum scenario to 4D gauge theory was first discussed (from very different points 
of view) by Luscher\cite{Luscher78} and Witten\cite{Witten78}. More recently, the same picture has emerged from
AdS/CFT duality \cite{Witten98}. 
The precise analogy between topological structure in 2D U(1) gauge theory and in 4D Yang-Mills theory was  
given by Luscher \cite{Luscher78}. He argued that the quantities which are analogous in the two cases 
are the Chern-Simons currents, i.e. the gauge-variant current whose divergence is the topological charge.
For a 2D U(1) gauge field $A_{\mu}$, this is
\begin{equation} 
j_{\mu}^{CS} = \epsilon_{\mu\nu}A^{\nu} \label{eq:2D}
\end{equation}
while for 4D SU(2) it is
\begin{equation}
   j_{\mu}^{CS} = \frac{1}{12\pi^2} \epsilon_{\mu\nu\rho\sigma}A^{\nu\rho\sigma} \label{eq:4D}
\end{equation}
where the 3-index tensor $A^{\nu\rho\sigma}$ is an {\it abelian} gauge field of the fourth kind
\begin{equation}
A^{\nu\rho\sigma} = - Tr\left\{B^{\nu}B^{\rho}B^{\sigma} + \frac{3}{2}B^{[\nu}\partial^{\rho}B^{\sigma]}\right\}\;\;,
\end{equation}
and B is the Yang-Mills field. In both 2D and 4D theories, a nonzero topological susceptibility implies the 
existence of ``hidden long-range order'' in the form of a massless $q^2=0$ pole in the correlator of two CS currents.
The analogy between (\ref{eq:2D}) and (\ref{eq:4D})
suggests a generalization of the k-vacuum scenario to 4D gauge theory. In the 2D case, the domain walls in spacetime are world lines of 
charged particles, which couple naturally
to the $A_{\mu}$ field via the usual Wilson line, $\int A_{\mu}dx^{\mu}$. 
For 4D Yang-Mills, the analogy suggests that $A_{\mu\nu\sigma}$ 
should couple to the 3-dimensional world volume of
a membrane or domain wall. Like the 2D Schwinger model, the k-vacua in 4D 
gluodynamics may be described as 
having an effective local value of $\theta_{eff}$ which differs from the 
value of $\theta$ in the Lagrangian by
integer multiples of $2\pi$. In the AdS/CFT view of $\theta$-dependence 
in 4D gauge theory\cite{Witten98}, the $\theta$ 
term arises from a 5-dimensional Chern-Simons term, with $\theta$ being 
given by the Wilson line of the
Ramond-Ramond U(1) field around the compactified fifth dimension. The fact that $\theta_{eff}$ changes by an integer 
multiple of $2\pi$ when crossing a domain wall is a Dirac-type quantization condition on the associated RR flux. 
 
\section{TC distributions in lattice CP(N-1)}

It is tempting to interpret the membrane-like regions of coherent topological charge observed in 4D QCD 
Monte Carlo configurations \cite{Horvath}
as a manifestation of the domain wall structure associated with k-vacua. Although it will take much additional
work to clearly establish such a connection, it is interesting to ask if a
similar phenomenon of long-range low-dimensional coherence occurs in Monte Carlo
configurations for the 2D CP(N-1) models. In fact we find precisely the behavior
that we would expect from Luscher's analogy between 2D U(1) and 4D YM. Typical configurations exhibit 
coherent regions of topological charge which form along one-dimensional ridges and
valleys which are generally long and narrow. A typical $Q(x)$ distribution for a CP(3)
configuration at $\beta = 3.0$ is shown in Fig. 1. We note that the $Q(x)$ plots
for CP(N-1) configurations bear a strong resemblance to plots of 2-dimensional slices
of 4D QCD configurations (c.f. Fig. 5 of Ref.\cite{Horvath}). 
Detailed studies to characterize the
size, shape, and scaling properties of these coherence regions are in progress and
will be discussed in a forthcoming paper.


\begin{figure}[t]
\vspace*{-0.20in}
\centerline{
  \includegraphics[width=7.0cm]{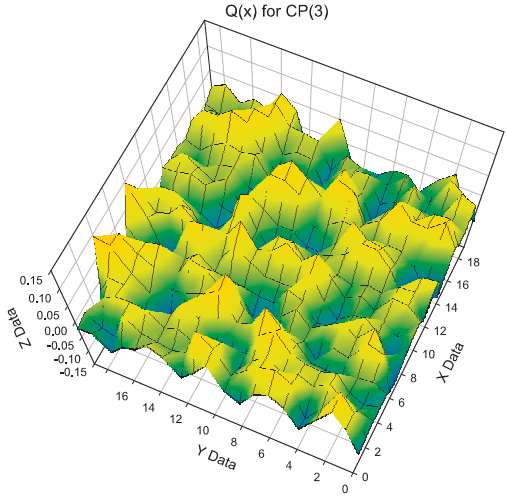}
           }
\vspace*{-.35in}
\caption{Topological charge distribution from overlap $Q$ for a 2D CP(3) configuration with $\beta=3.0$).}
\vspace*{-0.15in}
\end{figure}

The observation of long-range structure in TC distributions is made possible by using a 
definition of $Q(x)$ constructed from the overlap Dirac operator $D$,
i.e. $Q(x) = (1/2)Tr\gamma^5D(x,x)$. \cite{Hasenfratz}
In our calculations, we construct the overlap Dirac operator exactly, using the singular value decomposition routine
of LAPACK. The overlap definition of $Q$ turns out to be much smoother over short distances than the ultralocal
definition constructed from the plaquette phase. This ``chiral smoothing'' is clearly seen in the CP(N-1)
configurations, where the TC distributions from the plaquette phase exhibit no visible structure,
while the overlap $Q$ distributions show clear evidence of coherence. 
The 2-point correlators for the two definitions of $Q$ differ significantly
at short-distance,
e.g. the nearest-neighbor correlator for the plaquette phase is negative
(as required by reflection positivity), while for the overlap $Q$ it is positive
(allowed by nonultralocality).
Nevertheless, for moderate to large $\beta$,
the (integer valued) global topological charge obtained from the two 
definitions is almost always the same.

To construct a quantitative measure of coherence, we define a coherent structure as a
set of sites which all have the same sign of topological charge and are connected by
nearest-neighbor paths which remain within the structure.
Figure 2 shows the fraction of the lattice volume occupied by the $n$ largest structures for the
two definitions of $Q$. These results are from a large ensemble of $CP(9)$ configurations on a $50\times 50$ lattice with $\beta=0.8$ (correlation length$\approx 5$). Also shown for comparison is
the same plot for a set of random configurations. We see that the overlap definition of $Q$ exhibits a clear tendency toward coherence, e.g.
the typical largest structures are much larger than those in a random configuration. Surprisingly, the plaquette phase definition actually exhibits
{\it less} structure than the random case. This is an effect of the nearest-neighbor anticorrelation for the plaquette phase.

For a measure of the effective dimensionality (Hausdorff dimension) of the largest
structures, we start with each site on a structure and measure the number of other
sites $N(r)$ on the structure 
within a radius $r$. Fitting to $N(r)\propto r^d$, we obtain
$d=1.26(6)$, confirming the visual impression that these structure are approximately
one-dimensional. For comparison, we studied spin domains in the 2D Ising model just
above $T_c$, adjusted to give structures of the same volume as the CP(9)
configurations. The Hausdorff dimension of the Ising spin domains is found to be $d=1.86(5)$.

\begin{figure}[t]
\vspace*{-.00in}
\centerline{
  \includegraphics[width=6.0cm, angle=-90]{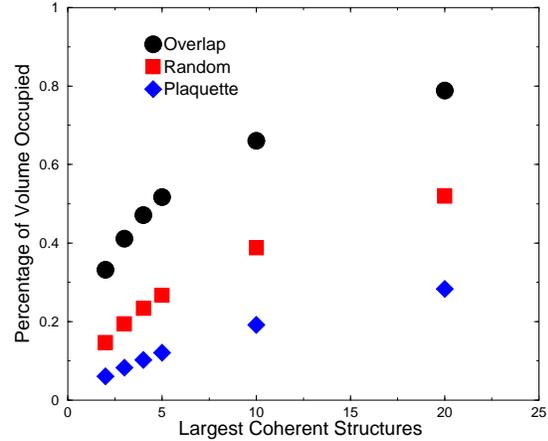}
           }
\vspace*{-.35in}
\caption{Percent of total volume occupied by the largest $n$ structures for overlap vs. plaquette definitions of topological 
charge. Also shown
is the same plot for random configurations.}
\vspace*{-0.15in}
\end{figure}

  \end{document}